\documentstyle[prd,eqsecnum,aps,epsf]{revtex}

\newcommand{\beq}{\begin{equation}}
\newcommand{\beqn}{\begin{eqnarray}} 
\newcommand{\eeq}{\end{equation}}
\newcommand{\eeqn}{\end{eqnarray}}
\newcommand{\beqa}{\begin{eqnarray}}
\newcommand{\eeqa}{\end{eqnarray}}

\newcommand{\gsim}{\mbox{\raisebox{-1.ex}{$\stackrel
     {\textstyle>}{\textstyle\sim}$}}}
\newcommand{\lsim}{\mbox{\raisebox{-1.ex}{$\stackrel
     {\textstyle<}{\textstyle \sim}$}}}

\newcommand{\be}{\begin{equation}}
\newcommand{\ee}{\end{equation}}
\newcommand{\bea}{\begin{eqnarray}}
\newcommand{\eea}{\end{eqnarray}}

\newcommand{\singlefig}[2]{
\begin{center}
\begin{minipage}{#1}
\epsfxsize=#1
\epsffile{#2}
\end{minipage}
\end{center}}
\newenvironment{figcaption}[2]{
 \vspace{0.3cm}
 \refstepcounter{figure}
 \label{#1}
 \begin{center}
 \begin{minipage}{#2}
 \begingroup \small FIG. \thefigure: }{
 \endgroup
 \end{minipage}
 \end{center}}

\begin{document}

\title{Fermion production from preheating-amplified metric perturbations}
\author{Bruce A. Bassett$^1$, Marco Peloso$^2$, Lorenzo Sorbo$^3$
and Shinji Tsujikawa$^4$ }
\address{$^1$ Relativity and Cosmology Group (RCG), University of
Portsmouth, Mercantile House, Portsmouth,  PO1 2EG, 
England\\[.3em]}
\address{$^2$ Physikalisches Institut, Universit{\"{a}}t Bonn Nussallee 12,
D-53115 Bonn, Germany}
\address{$^3$ SISSA/ISAS, via Beirut 2-4, I-34013 Trieste, Italy, 
and INFN, Sezione di Trieste, via Valerio 2, I-34127 Trieste, Italy}
\address{$^4$ Research Center for the Early Universe, 
University of Tokyo, Hongo, Bunkyo-ku, Tokyo 113-0033, Japan} 
\date{\today} 
 \maketitle
\begin{abstract}

We study gravitational creation of light fermions in the presence of 
classical scalar metric perturbations about a flat Friedmann-Lemaitre-
Robertson-Walker (FLRW) background. These perturbations can be large
during preheating, breaking the conformal flatness of the background
spacetime. We compute numerically the total number of particles generated
by the modes of the metric perturbations which have grown sufficiently to
become classical. In the absence of inhomogeneities massless fermions are
not gravitationally produced, and then this effect may be relevant for
abundance estimates of light gravitational relics.

\end{abstract}
\vskip 1pc 
\vskip 2pc

\section{Introduction}                            

It is well known that particles are created in an expanding Universe
\cite{BD80}. Pioneering work by Parker \cite{Parker} highlighted the
creation of nonconformally-invariant particles even in flat 
Friedmann-Lemaitre-Robertson-Walker (FLRW) Universes. The extension to
anisotropic Bianchi cosmologies by Zel'dovich and Starobinsky \cite{ZS},
showed that massless, conformally coupled, scalar particles are created
due to the breaking of conformal invariance which, in four dimensions, is
signaled by a non-vanishing Weyl tensor. In the presence of small,
inhomogeneous metric perturbations, Horowitz and Wald \cite{HW} obtained
the vacuum expectation value of the stress tensor for a conformally
invariant field and showed that particle production only occurs if
$\langle T_{\mu\nu}(x)\rangle$ is non-local (i.e., has non-zero
contributions with support on the past-null cone of the event $x$).

The number of particles created by inhomogeneities can be computed by a
perturbative evaluation of the $S$-matrix~\cite{Frieman,campos,CV}. In the
inhomogeneous case, the occupation number of produced particles is
composed by three parts, {\em viz}, the zeroth-order contribution due to
the homogeneous expansion, a first-order (in the  perturbation amplitude)
part arising from the interference between 0- and 2- particle states, and
a second-order contribution which comes from the interaction between
nonzero particle states. In the case of massive or nonconformally coupled
scalar fields, the last two terms (respectively linear and quadratic in
the metric perturbations) typically give a contribution small compared
with  the first one. However, in the massless and conformally coupled
cases,  the first two terms vanish in rigid, exactly FLRW backgrounds, and
only inhomogeneity contributes to the gravitational particle production.

Such theoretical studies were historically of interest both because they
focus on the interplay between quantum field theory and gravity and because
of their relevance to Misner's ``chaotic cosmology'' program \cite{misner}
in which arbitrary initial inhomogeneity and anisotropy were to be damped
to acceptable levels due to particle production. The non-renormalisability
of standard quantum gravity, the successes of string theory and the
dominance of cosmological inflation have since removed much of the
motivation for studies of particle production from inhomogeneity, though
research on non-equilibrium issues, generalized fluctuation-dissipation
relations and effective Einstein-Langevin equations still continues
\cite{CH}.

Inflationary cosmology seems to be less affected by the complex aspects of
gravity beyond the semi-classical realm. For inflation to start, a
homogeneous patch larger than the Hubble radius, $H^{-1}$, is needed which
may require some fine-tuning, especially for inflation at low energy 
scales. However, once inflation has begun, the no-hair conjecture
\cite{nohair}  should ensure that pre-existing inhomogeneities are driven
to  zero. The quantum metric perturbations generated during inflation can 
then
be tuned to have small  amplitudes $\sim 10^{-5}$ and unless one wants to
compute 4-point correlation functions which include graviton loops,
comparison with  the anisotropies in the Cosmic Microwave  Background
(CMB) is  straight-forward. 

However, nonperturbative production during the coherent oscillation
of the inflaton fields~\cite{TB,KLS,Boy,KT,PR}, dubbed
preheating~\cite{KLS}, can alter the simple picture of small amplitude 
metric  perturbations. Preheating therefore provides a particularly 
useful arena for examining some aspects of the boundary 
between semi-classical and quantum gravity.

Preheating can lead to the growth of metric perturbations, which
in turns is known to stimulate scalar particle creation
\cite{TN,massiveB,selfPE} and the production of seed magnetic fields
\cite{sting,FG,maroto,BPTV,DPTD}. The amplification of metric 
perturbations may also yield
runaway instabilities due to the negative specific heat of gravity, 
which leads to an interesting possibility of primordial black hole (PBH) 
formation \cite{PBHpre1,PBHpre2}. It can also amplify metric
perturbations on very large scales
\cite{sting,selfBV,selfFB,selfTBV,selfZBS,selfFK}. The growth of metric
perturbations, in the case where a PBH does not form, is limited by
backreaction effects which are explicitly nonlinear.

Hence on certain scales the Weyl tensor can actually be quite large. This
means that there can be significant particle production even if the field
is governed by a conformally invariant equation of motion, the classic
examples being photons, massless fermions and conformally coupled scalars.
In particular, if the effective mass of gravitational relics such as 
modulini and gravitini is small enough during reheating, then their  main
source of gravitational production will come from metric perturbations. 
Since stringent upper bounds hold for the abundances of  these species, it
is important to verify that creation from inhomogeneities  (if
sufficiently amplified at preheating) does not overcome this threshold.  
Our analysis suggests that this does not occur in the simplest models of 
chaotic inflation, at least when rescattering is neglected.

Particle creation via inhomogeneities may also be relevant for 
alternative scenarios such as the pre-big-bang model \cite{pbb}, where
homogeneous gravitational production of moduli and gravitinos is known to
be an issue \cite{pbb2} and the Universe goes through a high-curvature
phase where higher-order $\alpha'$ corrections are important.

The paper is organized as follows. In section II we review the formalism
for computing the number of produced fermions, while in section III we
discuss the evolution of metric perturbations during preheating in some
specific potentials. Sections IV and V present our numerical results and
conclusions respectively.

\section{Formalism for production of fermions} \label{forma}

In this section we review the formalism for the production of (Dirac)
fermionic quanta by cosmological inhomogeneities, summarizing the
calculations of~\cite{Frieman,campos,CV}. 
We consider  perturbations around the a FLRW background
\begin{equation}
g_{\mu\,\nu}=a^2\left(\eta\right)\,\left(\eta_{\mu\,\nu}
+h_{\mu\,\nu}\right)\,\,,
\label{gmunu}
\end{equation}
where $a(\eta)$ is the scale factor and $\eta$ the conformal time.
We will also use proper time, $t$, related to conformal time by
$d\eta= dt/a$ and choose to consider only scalar metric 
perturbations in the longitudinal
gauge~\cite{cosmo,pert} \footnote{Vector perturbations will induce 
magnetic fields via the induction equation \cite{TM} but we do not consider
them since inflation drives vector perturbations to zero. We also neglect
tensor (gravitational wave) modes since they are not strongly amplified 
at preheating.}.

At linear order minimally coupled scalar fields 
do not induce an anisotropic stress \cite{massiveB} 
and hence the metric perturbations 
are characterized by a single potential:
$h_{\mu\,\nu}=2\,\Phi\,\delta_{\mu\,\nu}\,$, {\em viz}:
\beqn
ds^2=a^2(1+2\Phi)d\eta^2
-a^2(1-2\Phi)\delta_{ij} dx^i dx^j\,\,.
\label{metric}
\eeqn
The action of a fermionic field $\psi$ with mass $m$ in 
the background~(\ref{metric})  is, to first order in $\Phi$,
\begin{eqnarray}
{\cal S} &=& \int\,d^4x\,a^3\,\left(1-2\,\Phi\right)\,\bar{\psi}\,\left\{ 
i \,\left[ \, \left( 1- \Phi \right)\, \gamma^0\, \partial_\eta + 
\frac{3}{2a} \frac{da}{d\eta}\,\left( 1- \Phi \right)\, 
\gamma^0-\frac{3}{2}\frac{d\Phi}{d\eta}\,\gamma^0\right. \right. 
\nonumber \\
&&\left.\left.+\left( 1+\Phi\right)\, \gamma^i\,\partial_i - \frac{3}{2} 
\,\gamma^i\,\partial_i\,\Phi \right]-a\,m\, \right\}\,\psi  \nonumber\\
&=& \int d^4x \: {\bar {\tilde{\psi}}} \left[  \left( i \gamma^0 \, 
\partial_\eta 
+ i \gamma^i \, \partial_i - a \, m\right) - 3 \, \Phi \, \gamma^0 \, 
\partial_\eta - \Phi \, \gamma^i \, \partial_i - 
\frac{3}{2}\frac{d\Phi}{d\eta}\, 
\gamma^0 - \frac{3}{2} \, \gamma^i \, \partial_i \Phi \right] \tilde{\psi} 
 \nonumber\\
&\equiv& 
\int d^4x \, \left[ {\cal L}_0 + {\cal L}_I \left( \Phi \right) 
\right]\,,
\label{feract}
\end{eqnarray}
where we have defined $\tilde{\psi} \equiv a^{3/2} \, \psi\,$ and 
${\cal L}_0$, ${\cal L}_I$ denote the homogeneous and interaction 
Lagrangians.  

The number density of produced particles is easily computed in the
interaction picture, where the evolution of the operators is just
determined by the homogeneous expansion of the Universe, and the states 
evolve according to the small inhomogeneities in the metric. The field 
$\tilde{\psi}$ can be decomposed in the standard way
\begin{equation} \label{decomp0}
\tilde{\psi} \left( x \right) = \int \frac{d^3 {\bf k}}{(2\pi )^{3/2}}
e^{i {\bf k} \cdot \mathbf{x}}\, \sum_r \left[ u_r \left( k, \eta \right) 
\, a_r \left( k \right) + v_r \left( k,\eta \right) b_r^\dagger \left( - k 
\right)\right] \,,
\end{equation}
where, as usual, the anti-commutation relations are
\begin{equation}
\left\{a_r(k), a_s^{\dagger}(k') \right\}=
\left\{b_r(k), b_s^{\dagger}(k') \right\}=
\delta^{(3)} \left( {\bf k}-{\bf k'} \right) \delta_{rs}.
\label{anti}
\end{equation}
The Fock space is built at the initial time $\eta_i$ starting from the 
vacuum state defined by
\begin{equation}
a_r \left( k \right) |0 \rangle = b_r \left( k \right) |0 \rangle = 0 \,\,.
\label{vacuum}
\end{equation}

For massless fermions (the conformally coupled case), the expansion of the 
Universe factors out of the free action $\int d^4 x \, {\cal L}_0$ 
and the spinors $u_r$ and $v_r$ evolve as in Minkowski spacetime. As a 
consequence, 
the operators ${a_r}^{(\dagger)} \,,\,{b_r}^{(\dagger)}$ are the physical 
annihilation (creation) operators at all times.  Introducing a mass term 
breaks the conformal symmetry, and the physical annihilation/creation 
operators are defined through the Bogolyubov transformations 
(see~\cite{prehferm} for details) %
\begin{eqnarray}
{\hat a}_r \left(k, \eta \right) &\equiv& \alpha_r \left(k, \eta 
\right) \, a_r \left( k \right) - {\beta_r}^* \left(k, \eta \right) \, 
{b_r}^\dagger \left( -\,k \right) \,\,, \nonumber\\
{{\hat b}_r}^\dagger \left(k, \eta \right) &\equiv& \beta_r
 \left( k, \eta \right) \, a_r \left( k \right) + {\alpha_r}^* \left(k, 
\eta 
\right) \, {b_r}^\dagger \left( -\,k \right) \,\,.
\end{eqnarray}
The two Bogolyubov coefficients have initial conditions $\alpha \left( 
\eta_i \right) = 1 \,,\, \beta \left( \eta_i \right) = 0\,$ and evolve 
according to~\cite{nps}
\begin{equation}
\alpha_r' = - \frac{m \, k \, a'}{2\, \omega^2} \, {\rm e}^{\,2 \, i 
\int^\eta \omega \, d  \eta}\,\beta_r \;\;,\;\; \beta_r' = \frac{m \, 
k \, a'}{2\, \omega^2} \, {\rm e}^{\,-\,2\,i \int^\eta \omega \, 
d \eta} \alpha_r \,\,,
\label{bogo}
\end{equation}
with $\omega \equiv \sqrt{k^2 + a^2 \, m^2}\,$. These equations preserve
the normalization $\vert \alpha_r \vert^2 + \vert \beta_r \vert^2 = 1\,$, which holds in the fermionic case.

In the interaction picture, the evolution of the initial vacuum (zero
particle) state $\vert 0 \rangle$ is determined by the interaction
Lagrangian ${\cal L}_I (\Phi)\,$. At first order in $\Phi$ we have
\begin{equation}
\vert \psi \rangle = \vert 0 \rangle + \frac{1}{2} 
\int d^3 k \, d^3 k' \,\vert k, r \,;\, k', s \rangle\,
\langle k, r \,;\, k', s \vert S \vert 0 \rangle \,\,,
\label{state}
\end{equation}
with the $S$-matrix element ($T$ stands for time ordering)
\begin{equation}
\langle k, r \,;\, k', s \vert S \vert 0 \rangle \equiv i \,T\,\langle k, 
r 
\,;\, k', s \vert {\cal L}_I \left( \Phi \right) \vert 0 \rangle \,\,.
\end{equation}

Rigorously speaking, the occupation number can be computed only in the
asymptotic future, $\eta_f$, with vanishing perturbations $\Phi \left( 
\eta_f \right) = 0\,$. Identical results are obtained for particles and
antiparticles, so we concentrate on the former. The expectation value
of the number operator ${\hat N} \equiv \left( 2 \, \pi \, a
\right)^{-\,3} \int d^3 p \, \hat{a}_r^\dagger \left(p, \eta_f
\right) \, \hat{a}_r \left(p, \eta_f \right)$ in the state $\vert
\psi \rangle$ is given by the sum of three terms, $N_0 + N_1 + N_2\,$,
which are of zeroth, first, and second order in $\Phi\,$, 
respectively ~\cite{Frieman,campos,CV},
\begin{eqnarray}
N_0 &=& \frac{V}{(2\pi a)^3}\int d^3 k \,
\langle 0 \vert \, {\hat a}_{r}^{\dagger}(k) \, {\hat a}_r (k) \, 
\vert 0\rangle =
\frac{V}{(2\pi a)^3} \int d^3 k \, \vert \beta_k \vert^2 \,\,,
\label{N0} \\
N_1 &=& - \, \frac{1}{(2\pi a)^3} \int d^3 k
\: {\rm Re} \, \left[ \alpha_r \left( k \right) \beta_r \left( k \right) 
\, \langle k, r \,;\, -\,k, r \vert S \vert 0 \rangle \right], 
\label{N1} \\
N_2 &=& \frac{1}{4\,(2\pi a)^3} \int d^3 k \, d^3 k' \, \vert 
\langle 0 \vert S \vert k, r  \,;\, k', s \rangle \vert^2 
\left[ \vert\alpha_r \left( k \right)\vert^2+\vert\beta_s \left( -\,k' 
\right)\vert^2+1 \right] \;\;\;.
\label{N2}
\end{eqnarray}
The zeroth-order term (\ref{N0}) is the well known expression arising
from  the homogeneous expansion of the Universe. $V$ denotes the volume of
the Universe at late times, when the perturbations can be neglected. The
first-order contribution (\ref{N1}) comes from the combined effect of
expansion and inhomogeneities. These two terms vanish for massless
fermions, which, as  we remarked, are conformally coupled to the FLRW
background (more  explicitly,  we notice that $\beta_r \left(k, \eta
\right) = 0$ in this case,  as can be seen from eqs.~(\ref{bogo})).

In the massless case, only the last term~(\ref{N2}) contributes to the 
particle production. Furthermore, in this limit $N_2$ acquires the 
particularly simple form~\cite{campos}
\begin{equation}
N_2 = \frac{1}{160 \, \pi \, a^3} \, \int \frac{d^4 p}{\left( 2 \, \pi 
\right)^4}\, \theta \left( p^0 \right) \, \theta \left( p^2 \right) \vert 
{\tilde C}^{abcd} \left( p \right) \vert^2 \,\,,
\end{equation}
in terms of the Fourier transform 
\begin{equation}
{\tilde C}^{abcd} \left( p \right) \equiv \int d^4 x \, {\rm e}^{i\,p\,x} 
\, C^{abcd} \left( x \right)
\end{equation}
of the Weyl tensor $C^{abcd}\,$.

For the metric~(\ref{metric}), one finds (see also~\cite{campos} for 
useful intermediate steps)
\begin{equation}
N_2 = \frac{1}{15\,\left( 2 \, \pi \right)^2 \, a^3} \, \int_0^\infty d
p_0 \int_{\vert {\bf p} \vert < p_0} d^3 {\bf p} \: \vert {\bf p} \vert^4
\: \left| \int_{\eta_i}^{\eta_f} d \eta \, {\rm e}^{i\,p_0\,\eta} \,
{\tilde \Phi} \, \left( {\bf p}, \eta \right) \right|^2 \,\,,
\label{npsi}
\end{equation}
where $\tilde{\Phi}$ is defined by
\begin{equation}
\Phi \left( \eta \,,\, {\bf x} \right) \equiv \int \frac{d^3 {\bf 
p}}{\left(2\,\pi\right)^{3/2}} \, {\tilde \Phi}
\left( {\bf p}, \, \eta \right) \, {\rm e}^{i \, {\bf p} \cdot {\bf x}} 
\;\;\;.
\label{esse}
\end{equation}
In these expressions, $\Phi$ should be regarded as a purely classical 
perturbation. The normalization as well as the classicality condition 
for $\Phi$ will be discussed in the subsequent sections.

\section{Evolution of scalar metric perturbations} \label{secpert}
\subsection{Linearized equations and analytic solutions for 
metric perturbations}

We now discuss the evolution of scalar perturbations $\Phi$ during
inflation and reheating in simple models of chaotic
inflation~\cite{chaos}, denoting the minimally coupled inflaton field by
$\phi$, with potential
\beqn
V =\frac12 \, m^2 \, \phi^2 \quad\quad {\rm or} \quad \quad V = \frac14 \, 
\lambda \, \phi^4\,.
\label{potential}
\eeqn
In subsection~\ref{twof} we will then discuss a model where two fields
are present.

For the potentials~(\ref{potential}), the COBE normalization
of the Cosmic Microwave Background Radiation~\cite{cobe} requires the
coupling constants be $m \sim 10^{-6} \, M_p$ for a massive inflaton
and $\lambda \sim 10^{-13}$ for the quartic case. Decomposing the
inflaton as $\phi (t, {\bf x}) \to \phi (t)+\delta \phi(t, {\bf x})$, the
background equations are given by
\begin{eqnarray}
  &&H^2 \equiv \left(\frac{\dot{a}}{a}\right)^2=
   \frac{8\,\pi}{3\,M_p^2}
   \left( \frac12 \dot{\phi}^2+ V \right) \,\,,
\label{hubble}\\
  &&\ddot{\phi}+3H\dot{\phi}+ \frac{d\,V}{d \, \phi} = 0\,\,,
\label{phi}
\end{eqnarray}
where dots denote derivatives with respect to physical time. In the
numerical simulations which we present in the next section, we will
implement the backreaction of field fluctuations within the Hartree
approximation. This corresponds to adding the contributions
\begin{eqnarray}
 \langle \delta\phi^2 \rangle=\frac{1}{2\pi^2} \int
k^2|\delta\phi_k|^2 dk,~~~
\langle \delta\dot{\phi}^2 \rangle=\frac{1}{2\pi^2} \int
k^2|\delta \dot{\phi}_k|^2 dk,~~~
\langle (\nabla\delta \phi)^2 \rangle=\frac{1}{2\pi^2} \int
k^4|\delta \phi_k|^2 dk,
\label{variance}
\end{eqnarray}
to Eqs.~(\ref{hubble}) and (\ref{phi}) (see
Refs.~\cite{KLS,Boy} for details). Note that this approach neglects
mode-mode coupling and rescattering effects \cite{KT}, which are 
important around the end of preheating,
and the backreaction effect of metric perturbations.
 The Fourier transformed, linearized Einstein
equations for field and metric perturbations in the
longitudinal gauge are \cite{pert}
\begin{eqnarray}
\dot{\Phi}_k + H\Phi_k = \frac{4\,\pi}{M_p^2}\, \dot{\phi}\delta \phi_k, 
\label{Phi1}
\end{eqnarray}
\begin{eqnarray}
3H\dot{\Phi}_k + \left( \frac{k^2}{a^2}+3H^2-
\frac{4\,\pi}{M_p^2} \, \dot{\phi}^2 \right) \Phi_k=
\frac{4\,\pi}{M_p^2} \, \left( \dot{\phi} \delta \dot{\phi}_k+
\frac{d^2V}{d \phi^2} \, \phi \delta\phi_k \right),
\label{Phi2}
\end{eqnarray}
%
\begin{eqnarray}
\delta\ddot{\phi}_k + 3H\delta\dot{\phi}_k+
\left( \frac{k^2}{a^2} + \frac{d^2V}{d \phi^2}
\right) \delta\phi_k=
2(\ddot{\phi}+3H\dot{\phi})\Phi_k+
4\dot{\phi}\dot{\Phi}_k.
\label{dphi}
\end{eqnarray}

The analytic form of the solutions of the above equations are known in
both the  limits of $k \to 0$ and $k \to \infty$ \cite{pert}. During
inflation, metric perturbations exhibit  adiabatic growth, $\Phi_k \simeq
c \dot{H}/H^2$,  after the first Hubble crossing ($k~\lsim~aH$). During
reheating, the super-Hubble modes $(k \ll aH)$ are nearly constant in the
single field case. In contrast, the solutions for the small-scale  modes
$(k \gg aH)$ can be described by  $\Phi_k \simeq \dot{\phi}\left(c_1e^{i
k\eta}+c_2e^{-i k\eta}\right)$, which shows adiabatic damping during
reheating.

The system of scalar metric fluctuations $\Phi_k$ and  inflaton
fluctuations $\delta \phi_k$ can also be described  in terms of the
Mukhanov-Sasaki variable $Q_k$ \cite{muksas}, defined by
\begin{equation}
Q_k  \equiv \delta \phi_k + \frac{\dot{\phi}}{H} \Phi_k \,\,,
\end{equation}
which satisfies the equation
\begin{equation}
\ddot{Q}_k + 3 \, H \, \dot{Q}_k + \left[
\frac{k^2}{a^2} +\frac{d^2V}{d \phi^2} + 
 \, \left( \frac{\dot{H}}{H} 
+ 3\, H \right)^\bullet \right] Q_k = 0 \,\,.
\label{eqq}
\end{equation}
The study of this equation is particularly convenient, since, contrary to 
the equation of motion for $\Phi_k$ alone, it is nonsingular during the 
oscillations of the inflaton field. The gravitational potential $\Phi_k$ 
is then related to $Q_k$ by
\begin{equation}
\frac{k^2}{a^2} \, \Phi_k = \frac{4\,\pi}{M_p^2} 
\frac{\dot{\phi}^2}{H} \, 
\left( \frac{H}{\dot{\phi}} \, Q_k \right)^\bullet \,\,.
\label{relqfi}
\end{equation}
During inflation, modes of cosmological interest initially (at $t =t_0$)
satisfy $k \gg a \, H$, and their equation of motion is that of a free
field in an expanding Universe. Quantization of this last quantity is thus
straightforward. In the initial vacuum state only positive frequency waves
are present (see~\cite{pert} for more details), so that we take as initial
conditions
\begin{equation}
Q_k \left( t_0 \right) = \frac{1}{a \left( t_0 \right) \, k^{1/2}} \, {\rm 
e}^{\,i \alpha_0} \;\;\;,\;\;\; \dot{Q}_k \left( t_0 \right) = 
\frac{-\,i\,k^{1/2}}{a^2\left( t_0 \right)} \, {\rm e}^{\,i \alpha_0}\,\,,
\label{eqqinitial}
\end{equation}
where $\alpha_0$ is an arbitrary phase.

As long as the modes remain much smaller than the Hubble scale 
they evolve as plane
waves. When their physical lengths grow to of order the Hubble radius 
($k \sim aH$), the variation of the frequency of $Q_k$ becomes important.
In fact, resonant amplification of some particular modes can occur during
the coherent inflaton oscillations at reheating \cite{TN}. This
phenomenon is mostly appreciated when the inflaton is nongravitationally
coupled to other fields \cite{selfBV} (see
subsection~\ref{twof}). However, some resonant amplification occurs also
in the  single-field self coupled case with potential $V=\lambda\phi^4/4$  
\cite{selfPE}, as we will discuss 
in the next subsection.

\subsection{Evolution of metric perturbations during reheating}
\subsubsection{$V=\frac12 m^2\phi^2$}

Let us first summarize the behavior of metric fluctuations during
reheating in the single field massive inflationary scenario.  During
reheating, from the time-averaged relation $\langle \dot{\phi}^2
\rangle=\langle m^2\phi^2 \rangle$ one finds $\phi \simeq
M_p/(\sqrt{3\pi}\,mt) \sin mt\,$. Neglecting in eq.~(\ref{eqq}) the
$H^2$ and $\dot{H}$ terms, which decrease as $\sim t^{-2}$ during
reheating, the rescaled variable $\tilde{Q}_k = a^{3/2} \, Q_k\,$ evolves
according to a Mathieu equation with time-dependent coefficients
\begin{eqnarray}
\frac{d^2}{dz^2}\tilde{Q}_k+(A_k-2q\cos 2z)
\tilde{Q}_k=0,
\label{Qk2}
\end{eqnarray}
where $z=mt/2\,$, and
\begin{eqnarray}
A_k=1+\frac{k^2}{(ma)^2},~~~~~
q=\frac{1}{\sqrt{2}z}.
\label{Aq}
\end{eqnarray}
The small $k$ modes ($k~\lsim~ma$) lie in the resonance band around
$A_k=1$. However, the cosmic expansion makes $q$ smaller than unity
already after one inflaton oscillation, and the resonance is not
efficient. As a consequence, modes in the long wavelength limit ($k \to
0$) grow as $\tilde{Q}_k \propto a^{3/2}$ \cite{KH,nata,FB}, which makes
$Q_k$ and $\Phi_k$ nearly constant during reheating (for the mode
$k~\sim~0.1m$ one has  $|k^{3/2}\Phi_k| \sim 10^{-5}$). On the contrary
the sub-Hubble modes ($k~\gsim~m)$ decrease  due to adiabatic damping
during reheating and are typically smaller than  the modes which already
crossed the Hubble scale  by more than one order of magnitude. This
behavior is shown for various wavelengths in Fig.~\ref{massivemet}.

\begin{figure}
\begin{center}
\singlefig{10cm}{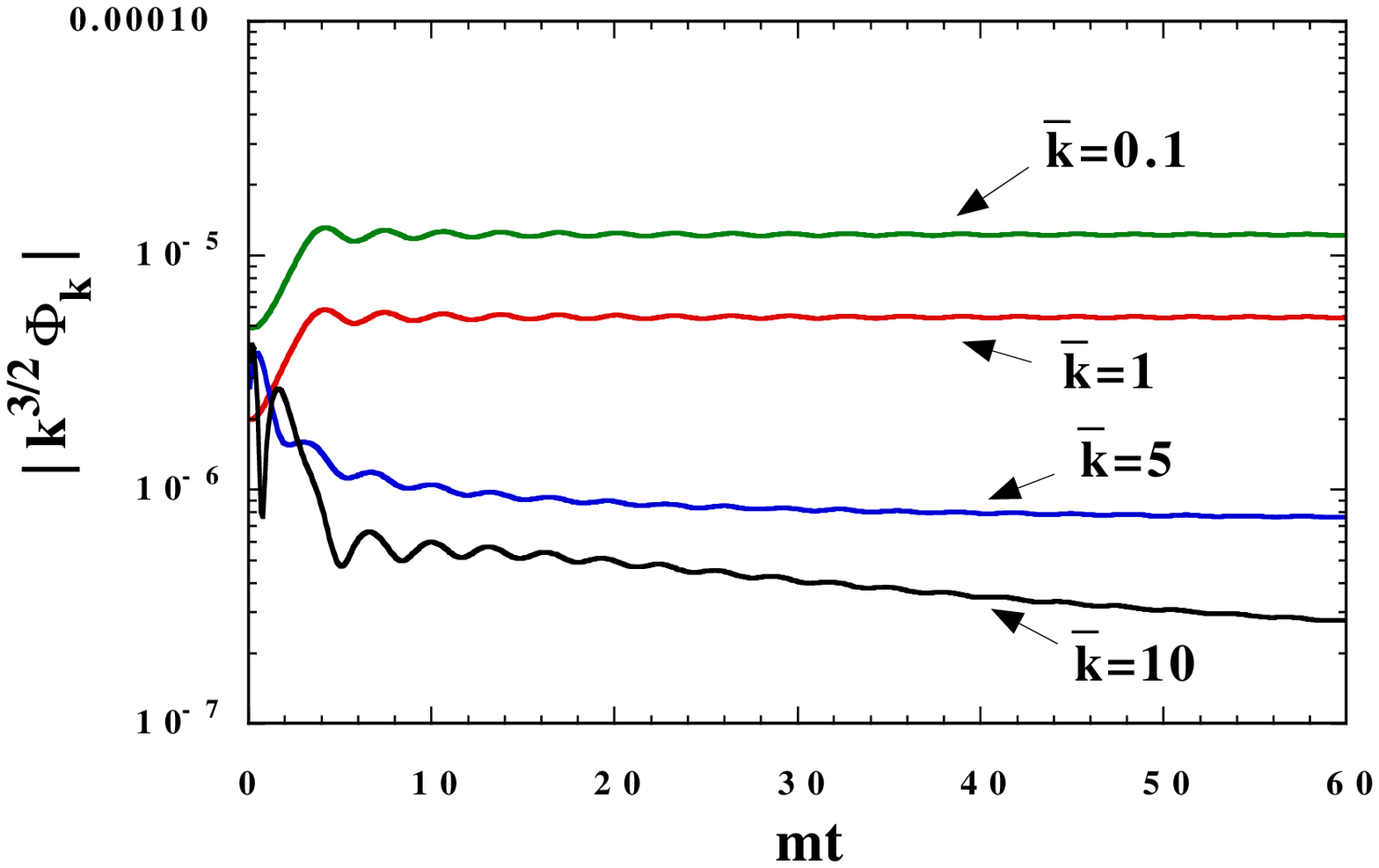}
\begin{figcaption}{massivemet}{10cm}
The evolution of metric perturbations for $\bar{k}\equiv
k/m=0.1, 1, 5, 10$ in the model $V=m^2\phi^2/2$ with inflaton mass
$m=10^{-6}M_p$. We start integrating with an initial value,
$\phi=0.5M_p$. The scale factor is normalized to unity at this time, and
$aH$ is always of order $m$ in the time range plotted. Modes with 
$k~\lsim~m$ exhibit adiabatic growth, and then (when
reheating starts, i.e., at $mt \approx 1.5$ in the plot)
$\Phi_k$ becomes nearly constant. Sub-Hubble modes ($k~\gsim~m$)
instead decrease during reheating due to the adiabatic damping of the
$\dot{\phi}$ term. 
\end{figcaption}
\end{center}
\end{figure}

%
\subsubsection{$V=\frac14 \lambda\phi^4$}

Let us next consider the single field massless chaotic 
inflation model, with 
potential $V(\phi)=\lambda\phi^4/4$. As shown 
in~\cite{Boy,Kaiser,GKLS}, the 
evolution of $\phi$ is described (on dropping the term $\frac{1}{a}
\frac{d^2a}{d\eta^2}$ from the 
equation of motion, eq.~(\ref{phi})) by
\begin{equation}
\phi \equiv \frac{\varphi}{a} = \frac{\phi_0}{a} \, {\rm cn} \left( x - 
x_0 \,,\, \frac{1}{\sqrt{2}} \right) \,\,,
\label{masslessev}
\end{equation}
where $x \equiv \sqrt{\lambda} \, \varphi_0 \, \eta$ is the dimensionless
conformal time, the suffix $0$ indicates the value of the quantities
at the beginning of reheating and ${\rm cn}(x-x_0, 1/\sqrt{2})$ 
is the elliptic cosine function.

Using the time-averaged relation $\langle \dot{\phi}^2 \rangle
=\lambda\phi^4$, one finds $(\dot{H}/{H})^{\cdot}  \propto
(\phi^2)^{\cdot} \propto a^{-3}$ and
$\dot{H} \propto a^{-4}\,$. Hence, the Hubble terms in 
Eq.~({\ref{eqq}) rapidly become negligible. 
Using $x$ and rescaling $\tilde{Q}_k = a \, Q_k$, 
Eq.~({\ref{eqq}) is reduced to
\begin{eqnarray}
\frac{d^2}{dx^2}\tilde{Q}_k+\left[\kappa^2+
3{\rm cn}^2\left(x, \frac{1}{\sqrt{2}}\right)\right]
\tilde{Q}_k=0 \,\,,
\label{lame}
\end{eqnarray}
where we have defined $\kappa^2 \equiv k^2/ \left( \lambda \, \varphi_0^2
\right)\,$, and set $x_0 = 0$ for simplicity. Eq.~({\ref{lame}) is a 
Lam\'e equation. From the analytical study of~\cite{GKLS} we deduce the
presence of one resonance band in the interval
\beqa 
\frac{3}{2}<\kappa^2<\sqrt{3} \,\,,
\label{single}
\eeqa
characterized by an exponential growth $\tilde{Q}_k \propto e^{\mu_k
x}\,$. The maximum value of the growth rate, $\mu_{\rm max}\approx
0.03598\,$, occurs at $\kappa^2 \approx 1.615$. This growth dominates 
over the dilution from the expansion of the Universe even for the 
unrescaled variable $Q_k\,$, and hence also for $\Phi_k\,$. The growth of
$\Phi_k$ continues until the backreaction of inflaton fluctuations shuts
off the resonance. The maximal value of the metric perturbation  is found
numerically to be $\vert k^{3/2} \, \Phi_k \vert \simeq 5 \times
10^{-\,5}\,$. Modes outside the interval~(\ref{single}) do not exhibit
nonadiabatic growth, unless rescattering effects are taken into account
\cite{KT,selfPE}.

\begin{figure}
\begin{center}
\singlefig{10cm}{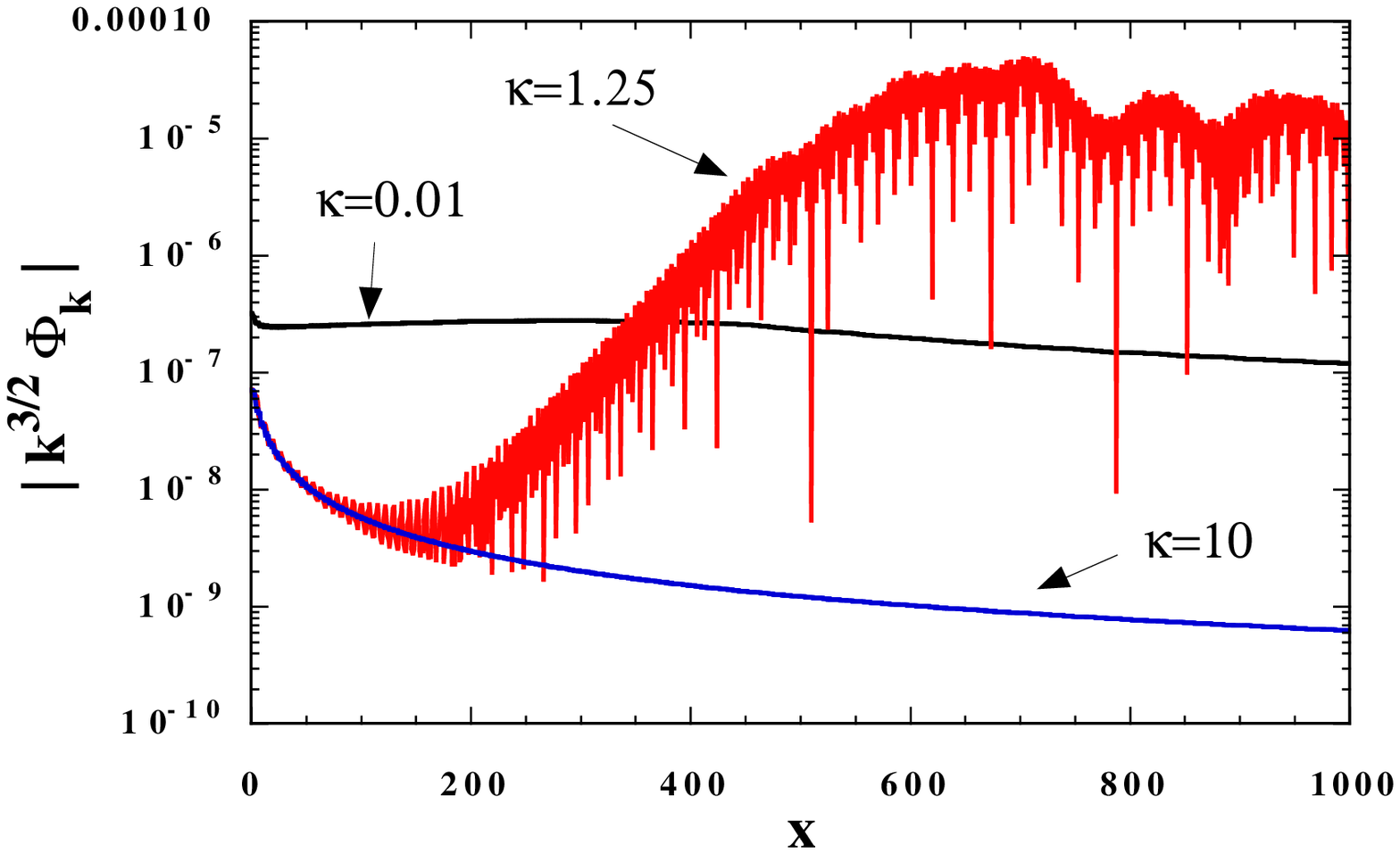}
\begin{figcaption}{selfmet}{10cm}
The evolution of $\Phi_k$ for $\kappa=0.01, 1.25, 10$ in the single field
potential $V=\lambda\phi^4/4$ and $\lambda=10^{-13}$. The initial 
values of the numerical evolution are $\phi=0.1M_p\,, a=1\,$. We 
include the backreaction effect of produced particles via the Hartree 
approximation (see the text for details). Notice the enhancement of the 
mode $\kappa=1.25$ in the sub-Hubble resonance band~(\ref{single}).
\end{figcaption}
\end{center}
\end{figure}

%
\subsubsection{$V=\frac14 \lambda\phi^4+\frac12 
g^2\phi^2\chi^2$} \label{twof}

We now discuss the situation in which a massless inflaton $\phi$ is
nongravitationally coupled to another scalar field $\chi$, with
conformally invariant potential $V=\lambda\phi^4/4+g^2\phi^2\chi^2/2$. 
This model was studied in detail, in the absence of
metric perturbations, in Refs.~\cite{Kaiser,GKLS}, showing the
presence of  an infinite number of strong resonance bands for $\chi$ as a
function of
\begin{equation}
R \equiv g^2/\lambda \,\,.
\end{equation}
The conformal invariance allows exact Floquet theory to be used to prove
analytically that metric fluctuations can exhibit parametric amplification
\cite{selfBV}. This makes the model a very useful testing ground for a
number of issues \cite{selfFB,selfTBV,selfZBS,selfFK} including the
possible formation of primordial black holes in some parameter space
regions \cite{PBHpre2}. Here we briefly summarize those of the above
results which are relevant for the calculations of the next section.

As before, it is convenient to introduce a Mukhanov-Sasaki variable for
the each field, $Q_k^{(i)}\equiv \delta \varphi_k^{(i)} + \left(
\dot{\varphi}^{(i)} / H \right)\Phi_k$, $i = \phi,\chi$ which satisfy a
simple generalization of eq.~(\ref{eqq}) \cite{selfBV,selfFB}.

Using eq. (\ref{masslessev}) and rescaling all variables as 
${\tilde F} \equiv a \, F\,$, the equation for
$\tilde{Q}_k^{\chi}$ reads \cite{selfBV}
\begin{eqnarray}
\frac{d^2}{dx^2}\tilde{Q}_k^{\chi}+\left[\kappa^2+
\frac{g^2}{\lambda}{\rm cn}^2\left(x, \frac{1}{\sqrt{2}}\right)\right]
\tilde{Q}_k^{\chi}=-2\frac{g^2}{\lambda \phi_0} 
{\rm cn}^2\left(x, \frac{1}{\sqrt{2}}\right) \tilde{\chi} 
\tilde{Q}_k^{\phi}+M_{\phi\chi}\tilde{Q}_k^{\phi}
+M_{\chi\chi}\tilde{Q}_k^{\chi} \,\,,
\label{lame2}
\end{eqnarray}
where
\beqn
M_{\varphi_1\varphi_2}=
\frac{8\pi}{a^2M_{p}^2} \left[\frac{1}{aH}
\left(\tilde{\varphi}_1'-\frac{a'}{a}\tilde{\varphi}_1
\right) \left(\tilde{\varphi}_2'-\frac{a'}{a}\tilde{\varphi}_2
\right) \right]' \,\,.
\label{M}
\eeqn
Here prime denotes a derivative with respect to $\eta$.
In the early stages of preheating, as long as $\chi$ fluctuations are
small relative to $\phi\,$, one can neglect the right hand side of
eq.~(\ref{lame2}), which then reduces to the generalized Lam\'e equation
studied in~\cite{Kaiser,GKLS} (in the following, we will apply the general
analytical results for the Lam\'e equation given in~\cite{GKLS}). Of
particular cosmological interest are the longwave modes on the scales
proben by the current CMB experiments. It can be shown~\cite{GKLS} that
the resonance bands extends to very longwave modes ($\kappa \to 0$) only
when the ratio $R$ is in the range
\begin{eqnarray}
n \, \left( 2n-1 \right) < R < n \, \left( 2n+1 \right) 
\,\,,
\label{range}
\end{eqnarray}
where $n$ is a positive integer. In each of these intervals, the growth
rate $\mu_k$ for $\tilde{Q}_k^{\chi}$ can be analytically found as
\begin{eqnarray}
\mu_k=\frac{2}{T} {\rm ln}
\left(\sqrt{1+e^{-\pi \epsilon^2}}+e^{-\pi \epsilon^2/2}\right),
\label{rate}
\end{eqnarray}
where $\epsilon \equiv \sqrt{2/R}\,\kappa^2$ and $T \simeq 7.416$ is the
period of inflaton oscillations. Then we have maximal growth when $R$ is
exactly $2\,n^2$ and $\kappa=0$, with large characteristic exponent
$\mu_{\rm max}=(2/T) {\rm ln} (\sqrt{2}+1)\simeq 0.2377\,$.

It is important to recognize that the exponential growth of
$\tilde{Q}_k^{\chi}$ just discussed does not necessarily translate into a
parametric amplification of the gravitational potential $\Phi_k\,$. In the
two-field case, Eq.~(\ref{Phi1}) is modified to
\begin{eqnarray}
\dot{\Phi}_k + H\Phi_k = \frac{4\,\pi}{M_p^2}
(\dot{\phi}\delta \phi_k+\dot{\chi}\delta \chi_k) \,\,.
\label{Phim}
\end{eqnarray}
When $\chi$ and $\delta \chi_k$ are vanishingly small relative to  $\phi$
and $\delta \phi_k$, the evolution of $\Phi_k$ is similar to  the single
field case. In fact, when $R \gg 1$ the homogeneous mode $\chi$ as well as
the longwave $\delta\chi_k$ modes ($k \to 0$) are exponentially suppressed
during  inflation \cite{selfBV,selfFB,selfTBV,selfZBS} due to the large
effective mass $g \, \phi$ relative to the Hubble parameter (see also
Ref.~\cite{suppression,liddle,shinji}). This can be explicitly seen by
considering the equation of motion for  $\delta \chi_k\,$, which for
super-Hubble modes reads
\begin{equation}
\ddot{\delta \chi_k} \, + 3 \, H \, \dot{\delta \chi_k} + g^2 \, \phi^2 \, 
\delta \chi_k \simeq 0 \,.
\label{delchisup}
\end{equation}
Hence, during inflation, the amplitudes of the long-waves modes
evolve as \cite{selfBV}
$|\delta\chi_k| \propto a^{- \left( 3/2-\nu \right)}\,$, with  
$\nu = {\rm Re}\left[9/4 - 3 \, R \,M_p^2 / 
\left( 2 \, \pi \, \phi^2 \right) \right]^{1/2}$.
This confirms that the inflationary suppression of the long-waves modes
becomes more  important as $R$ increases. In addition, considering
backreaction in the Hartree approximation, one can show~\cite{selfZBS} that
for $R~\gsim~8$ the growth of  field perturbations shuts off the resonance
before the super-Hubble metric perturbations ($k \to 0$) begin to be
amplified. Hence, for large $R\,$, the evolution of the gravitational
potential $\Phi_k$ on the scales probed by CMB experiments does not
significantly differ from the ``standard'' one described in~\cite{pert}
unless rescattering is taken into account. 
For these modes, 
parametric amplification is expected to
be effective in the first resonance band ~(\ref{range}) even when
backreaction is considered~\cite{selfBV,selfFB,selfTBV,selfZBS}.

\begin{figure}
\begin{center}
\singlefig{10cm}{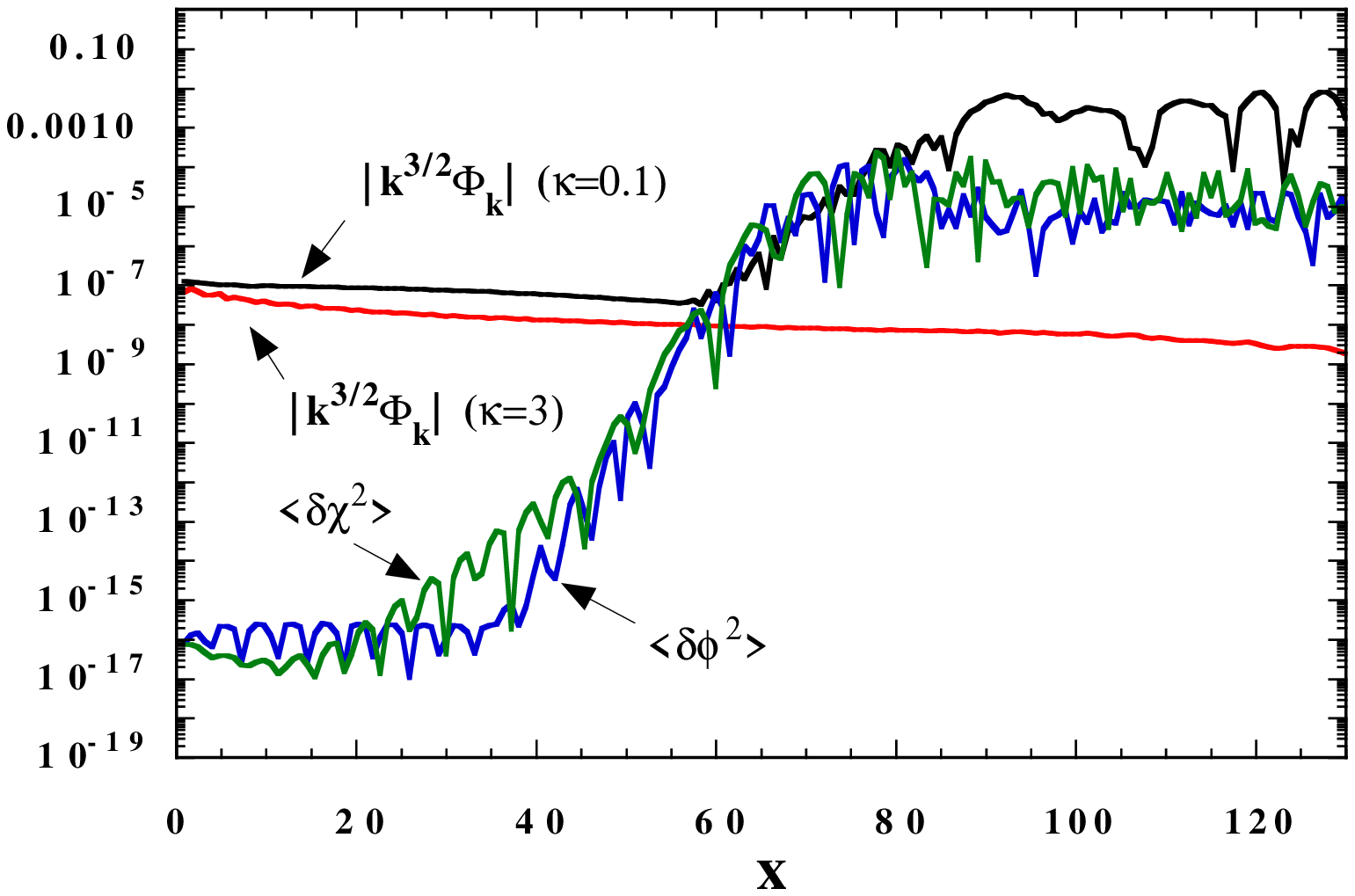}
\begin{figcaption}{selfmetmulti}{10cm}
The evolution of metric perturbations $\Phi_k$ and field variances 
$\langle\delta\chi^2\rangle$ and $\langle\delta\phi^2\rangle$ for 
$\kappa=0.1, 3$ in the potential $V=\lambda\phi^4/4+g^2\phi^2\chi^2/2$  
with $g^2/\lambda=2$ and
$\lambda=10^{-13}$.  We start integrating with initial values
$\phi=0.1M_p$  and $\chi=5 \times 10^{-6}M_p$,
corresponding to $\chi=10^{-3}M_p$  at $\phi=4M_p$ 
(i.e., 55 e-folds before the end of inflation). 
The final results are weakly
dependent on the choice of the initial $\chi$ as long as
$10^{-6}M_p<\chi<M_p$ at $\phi=4M_p$ \cite{selfZBS,PBHpre2}. 
\end{figcaption}
\end{center}
\end{figure}

In Fig.~\ref{selfmetmulti} we plot the evolution of $\Phi_k$ for the two
modes $\kappa=0.1$ and $\kappa=3$ in the $R=2$ case. We find that metric
perturbations begin to grow after the $\delta\chi_k$ fluctuations are
sufficiently amplified. The growth of $\Phi_k$ ends around $x=90$ where 
the backreaction terminates the resonance, after which perturbations reach
the plateau region  found in Fig.~\ref{selfmetmulti}. 

While only long-wave modes are necessary when comparing with 
CMB measurements, particle production is also induced by modes $\Phi_k$ 
with
larger momentum. The sub-Hubble modes ($k~\gsim~aH$) at the beginning of
reheating are not severely suppressed during inflation, and therefore the
modes $\delta \chi_k$ on those scales can be amplified to large values
during preheating. However, from $m_\chi = g \, \langle \phi \rangle$ we
see that large values of $R=g^2/\lambda$ results in a suppression of
the homogeneous mode $\chi$ during inflation. This also suppresses 
amplification of $\Phi$ from the rhs of equation~(\ref{Phim}).
Indeed, for $R \gg 1$ the dominant effect for the source of metric
perturbations is of second order in nonadiabatic (isocurvature) field
perturbations \cite{liddle}. To estimate this growth, we consider the
curvature perturbation on uniform-density hypersurfaces \cite{BST,adient},
denoted by $\zeta\,$. This quantity is related to the gravitational 
potential $\Phi$ via \cite{adient}
\begin{eqnarray}
-\zeta={\cal R}+\frac{2\rho}{3(\rho+p)}
\left(\frac{k}{aH}\right)^2 \Phi,
\label{zeta}
\end{eqnarray}
where ${\cal R}\equiv \Phi-(H/\dot{H})(\dot{\Phi}+H\Phi)$ is the comoving
curvature perturbation, and $p$ and $\rho$ are the pressure and the energy
density, respectively. 
The variation of $\zeta$ occurs on large scales in the presence of a 
nonadiabatic contribution to the pressure perturbation, $\delta p_{\rm 
nad}= \dot{p}\left(\delta p/\dot{p}-\delta\rho/\dot{\rho}\right)$.  One can 
estimate the second order contribution to $\zeta$, denoted $\zeta_{(2)}$, 
as \cite{liddle} 
\begin{eqnarray}
\zeta_{(2)}=\frac{3H}{\dot{\rho}} \int
\delta p_{\rm nad}Hdt \simeq
\frac{1}{\dot{\phi}^2} \int \left(1+\frac{2\lambda\phi^3}
{3H\dot{\phi}}\right) g^2\phi^2 \delta\chi^2 H dt,
\label{zetanon}
\end{eqnarray}
where we used  the time averaged relation,
$\delta\dot{\chi}^2 \simeq m_{\chi}^2 \delta\chi^2 \simeq
g^2\phi^2\delta\chi^2$.
We find that the second order term, $g^2\phi^2 \delta\chi^2$,
can induce a variation of $\zeta$, while the first order term
including the homogeneous $\chi$ is vanishingly small for $R \gg 1$.
{}From eq.~(\ref{zetanon})  we find
\begin{eqnarray}
\left|k^{3/2}\zeta_{(2)}(k) \right|
\simeq \frac{\sqrt{2}\pi}{\dot{\phi}^2} \int
 \left(1+\frac{2\lambda\phi^3}
{3H\dot{\phi}}\right) g^2\phi^2 \sqrt{P_{\delta\chi_k^2}} H dt \,,
\label{preper2}
\end{eqnarray}
where $P_{\delta\chi_k^2}$ is defined by $P_{\delta\chi_k^2} \equiv
\frac{k^3}{2\pi^2} \langle |\delta \chi_k^2|^2 \rangle$.
Following Ref.~\cite{liddle}, the power spectrum of the $\delta\chi_k$
fluctuation is estimated as 
\begin{eqnarray}
P_{\delta\chi_k } \equiv
\frac{k^3}{2\pi^2} \langle |\delta \chi_k|^2 \rangle
\simeq 
\frac{H_0}{g\phi_0} \left( \frac{H_0}{2\pi}\right)^2
\left( \frac{\kappa}{\kappa_0}\right)^3 
\frac{e^{2\mu_k x}}{a^2} \,.
\label{Ppre}
\end{eqnarray}
Here $\kappa_0 \simeq 0.15$ is the mode corresponding
to the horizon size at $\phi=0.1M_p$.
{}From eq.~(\ref{Ppre}) one finds
\begin{eqnarray}
P_{\delta\chi_k^2} \equiv
\frac{k^3}{2\pi^2} \langle |\delta \chi_k^2|^2 \rangle
\simeq
\frac{k^3}{2\pi} \int_0^{k_c} \frac{P_{\delta\chi}
(|{\bf k'}|) P_{\delta\chi}(|{\bf k-k'}|) }{|{\bf k'}|^3
|{\bf k-k'}|^3} d^3 {\bf k'}
\simeq
\left(\frac{H_0}{g\phi_0}\right)^2 \left( \frac{H_0}{2\pi}\right)^4
\left( \frac{\kappa}{\kappa_0}\right)^3 \frac{M(\kappa, x)}{a^4},
\label{Pdeltachi2}
\end{eqnarray}
where
\begin{eqnarray}
M(\kappa, x) &\equiv& \left(\frac{\sqrt{\lambda}\phi_0}{\kappa_0}
\right)^3 \int_0^{\kappa_c} d\kappa' \int_0^{\pi} d\theta
e^{2(\mu_{\kappa'}+\mu_{\kappa-\kappa'})x} \kappa'^2
\sin \theta \nonumber \\
&\simeq&
2 \left(\frac{\sqrt{\lambda}\phi_0}{\kappa_0}
\right)^3 \int_0^{\kappa_c} d\kappa' e^{4\mu_{\kappa'}x}
\kappa'^2 +{\cal O} (\kappa^2).
\label{Meq}
\end{eqnarray}
Note that we expanded  the term $\mu_{\kappa-\kappa'}$ around 
$\mu_{\kappa'}$ in eq.~(\ref{Meq}). Due the $\kappa^3$ factor in
eq.~(\ref{Pdeltachi2}), the spectrum $P_{\delta\chi_k^2}$ vanishes  in the
large scale limit ($k \to 0$), which implies that also second order
effects in field perturbations do not induce large metric perturbations on
the scales probed by CMB measurements~\cite{liddle}. In contrast, on
sub-Hubble scales at the beginning of preheating ($\kappa \sim \kappa_0$),
$P_{\delta\chi_k^2}$ is nonvanishing. In this case parametric excitation
of the $\chi$ fluctuation can lead to a growth of $\zeta$ and $\Phi$.

\begin{figure}
\begin{center}
\singlefig{10cm}{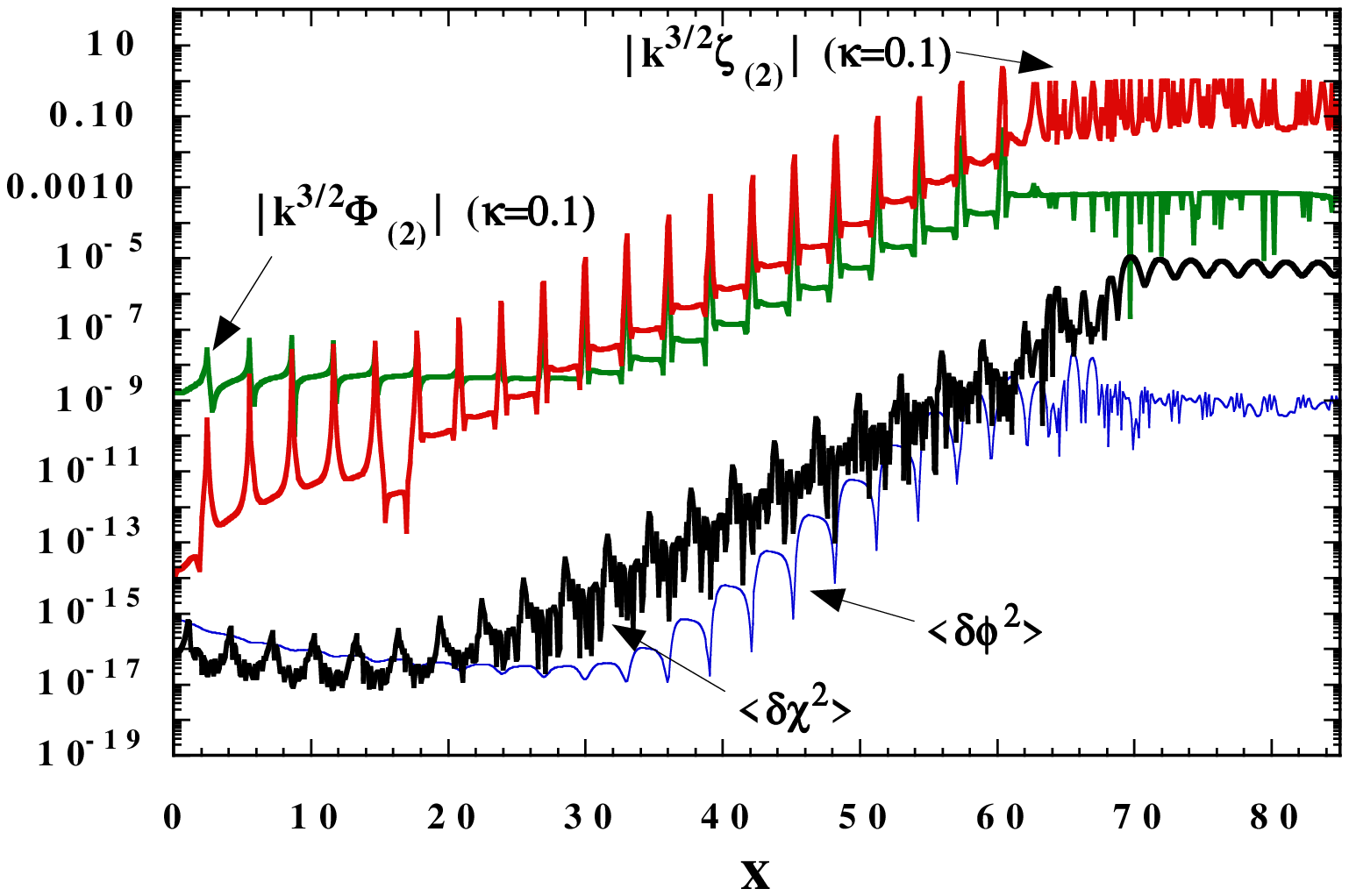}
\begin{figcaption}{second}{10cm}
The evolution of the gravitational potential and the
curvature perturbation for  $\kappa=0.1$  in the model of $V=
\lambda\phi^4/4+g^2\phi^2\chi^2/2$   with $g^2/\lambda=5000$ and
$\lambda=10^{-13}$.  Note that we include the second order effect in field
perturbations. We find that metric perturbations are enhanced on
sub-Hubble scales. We also plot the evolution of the field variance, 
$\langle\delta\chi^2\rangle$ and $\langle\delta\phi^2\rangle$, within
the Hartree approximation.
\end{figcaption}
\end{center}
\end{figure}

As an example, we plot in Fig.~\ref{second} the evolution of 
$|k^{3/2} \Phi_{(2)}|$ and $|k^{3/2}\zeta_{(2)}|$
for $\kappa=0.1$ in the case of $g^2/\lambda=5000$. 
$|k^{3/2}\Phi_{(2)}|$ is evaluated by making use of eqs.~(\ref{zeta}) 
and (\ref{preper2}).  Note that the final field variances are smaller than 
in the $g^2/\lambda=2$ case, due to the stronger backreaction effect (c.f.  
Fig.~\ref{selfmetmulti}).  The growth of nonadiabatic perturbations 
continues until backreaction ends the resonance.  Although the final value 
of $|k^{3/2}\Phi_{(2)}|$ is somewhat smaller than that of 
$|k^{3/2}\Phi_k|$ in the $g^2/\lambda=2$ case, the fermion 
number density (\ref{npsi}) can be larger due to the contribution of higher 
momentum modes, as we will see in the next section.

\section{Numerical results for the production of fermions}
\subsection{Quantum to classical transition}

The results of section~\ref{forma} refer to particle production by a
classical external source. On the contrary, the relation (\ref{relqfi}) of
the previous section relates the modes of the decompositions
\begin{equation}
\Phi (\eta,\,{\bf x})=\frac{1}{\sqrt{2}}\,\int\,
\frac{d^3{\bf k}}{\left(2\,\pi\right)^{3/2}}\left[\Phi_k
\left(\eta\right)\,e^{i{\bf 
{k \cdot x}}}\,a(k)+\Phi^*_k\left(\eta\right)\,
e^{-i{\bf {k\cdot x}}}\,a^\dagger(k) \right]\,\,,
\label{deffik}
\end{equation}
\begin{equation}\label{defqk}
Q^{(i)}(\eta,\,{\bf x})=\frac{1}{\sqrt{2}}\,\int\,\frac{d^3{\bf 
k}}{\left(2\,\pi\right)^{3/2}}\left[Q_k^{(i)}\left(\eta\right)\,
e^{i{\bf {k\cdot x}}}\,a(k)+Q^*_k{}^{(i)}
\left(\eta\right)\,e^{-i{\bf {k \cdot x}}}\,
a^\dagger (k)\right]\,\,,
\end{equation}
to the annihilation and creation operators $a(k)$ and $a^\dagger(k)$. As is
well known, metric perturbations arise from quantum fluctuations during
inflation and some of them eventually undergo a transition to
classicality. A consistent treatment of fermion production requires the
integral (\ref{npsi}) to be limited to those modes of $\Phi$ that have
grown enough to become classical.

A simple criterion for classicality is given in Ref.~\cite{PS}.
Classicality is expected to be a good approximation when the following
condition is satisfied:
\begin{equation}
\Delta_k^{(i)} \equiv \left| Q_k ^{(i)} \, a  \right| 
\left| ( Q_k^{(i)} \, a )' 
\right| \gg 1 \,\, ~~~~~~~~i = \phi,\chi
\label{class}
\end{equation}
The quantity $\Delta_k^{(i)}$ is related to the uncertainty in the
determination of $\langle \Delta Q^{(i)} \, \Delta \Pi^{(i)} \rangle\,$,
where $\Pi^{(i)}$ is the conjugate momentum of $Q^{(i)}\,$. In the case of
pure vacuum fluctuations, the equality $\Delta_k ^{(i)} = 1$ holds, in
accordance with the rules of quantum mechanics. Notice that this is the
case for the initial states~(\ref{eqqinitial}). A much bigger
uncertainty~(\ref{class}) is associated to large fluctuations of classical 
nature.

\begin{figure}
\begin{center}
\singlefig{10cm}{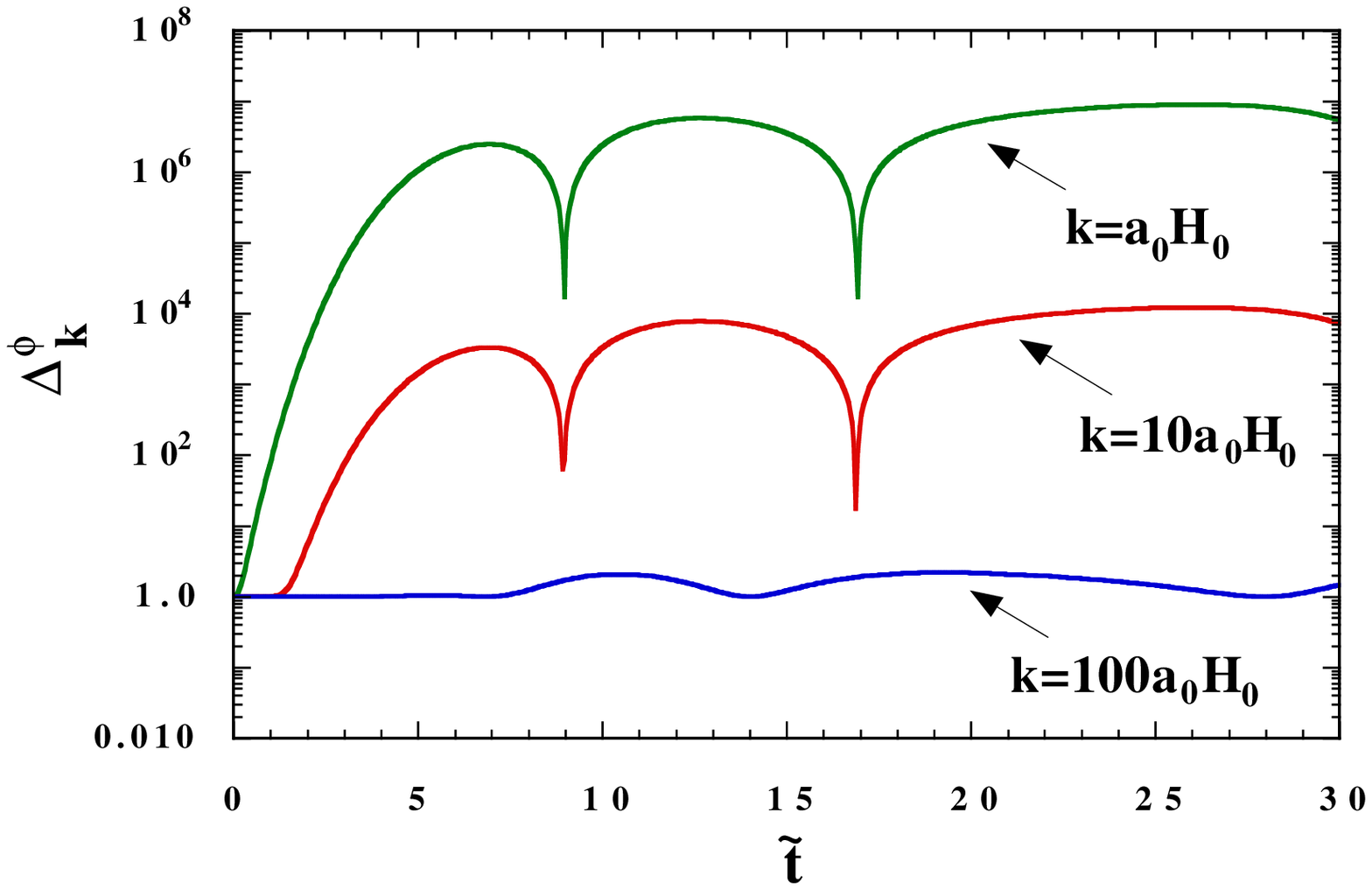}
\begin{figcaption}{deltak}{10cm}
The quantum to classical transition as signaled by the evolution of  the
quantity $\Delta_k^{\phi}$ during inflation and subsequent reheating stage
in the single field model,  $V=\lambda\phi^4/4$. The mode $k=a_0H_0$
leaves the Hubble scale  at $\phi=1.5M_p$ (i.e., at the initial time
plotted). $\Delta_k^{\phi}$ grows significantly from  unity after Hubble
radius crossing. In this figure the end of inflation corresponds to 
$\tilde{t}\equiv \sqrt{\lambda}\phi_0 t \approx 10$.
\end{figcaption}
\end{center}
\end{figure}

\begin{figure}
\begin{center}
\singlefig{10cm}{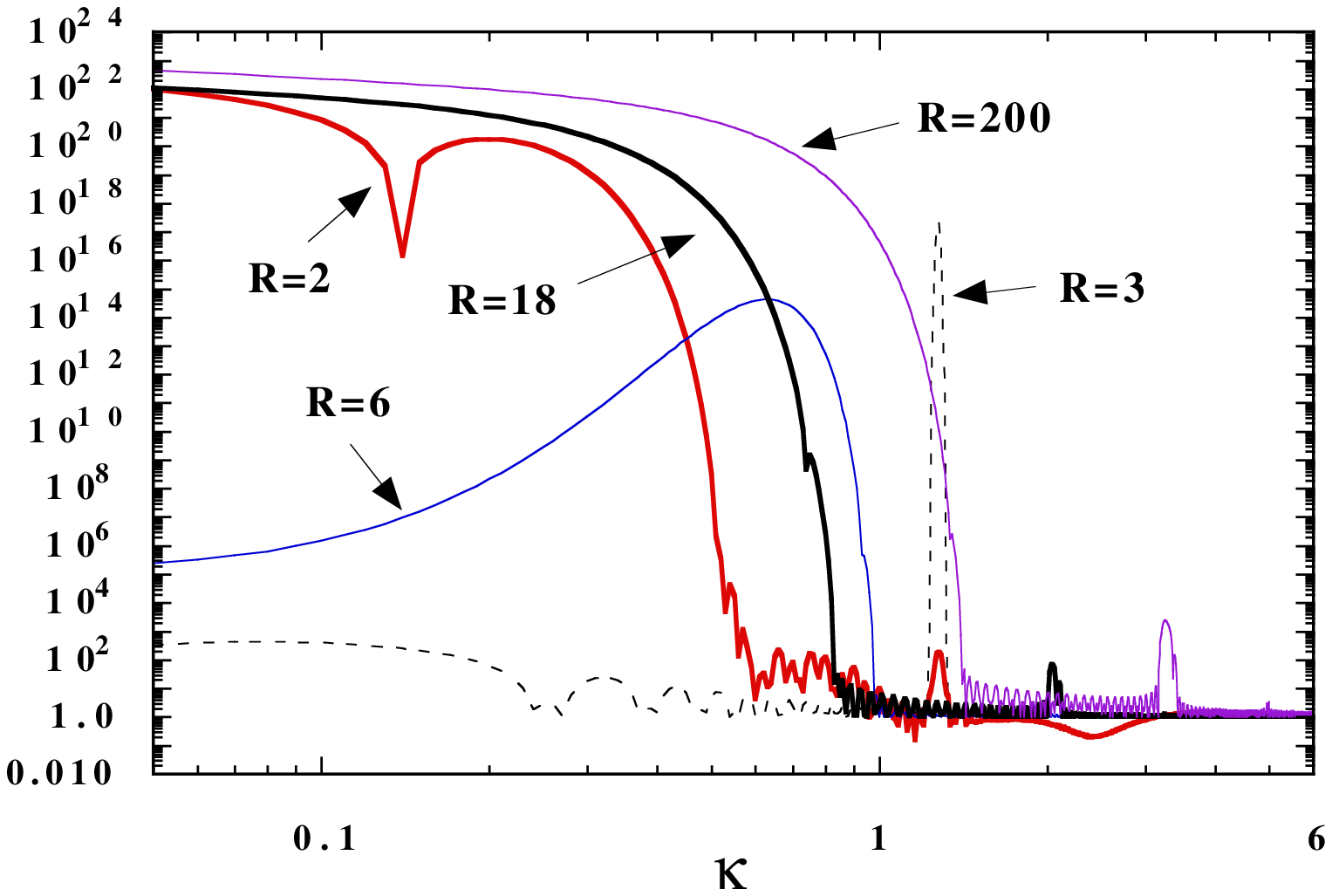}
\begin{figcaption}{deltamulti}{10cm}
The spectra of $\Delta_k^{\chi}$ when the variance 
$\langle \delta \chi^2 \rangle$ reaches its maximum value
in the two-field model of preheating, 
$V=\lambda\phi^4/4+g^2\phi^2\chi^2/2$,
for different values of $R=g^2/\lambda$.
The small $k$ modes ($\kappa<0.5$) are
sufficiently enhanced by parametric resonance 
for $R=2n^2$, which makes the perturbations
highly classical. In the case of $R<20$ we find that 
$\Delta_k^{\chi}$ is very close to unity
for the modes, $\kappa~\gsim~3$.
Therefore the cut-off $\kappa_c=3$ is justified in evaluating 
the occupation number of fermions for  $R<20$.
\end{figcaption}
\end{center}
\end{figure}

In Fig.~\ref{deltak} the time evolution of $\Delta_k^{\phi}$ is  shown in
the model $V = \lambda \, \phi^4/4$, for some particular values of $k\,$.
We notice that $\Delta_k^{\phi}$ begins to increase around the region
where a mode leaves the Hubble scale during inflation. This is expected to
occur quite generically, irrespective of the specific form of the inflaton
potential. On the contrary, modes which are always inside the Hubble scale
typically conserve their quantum nature ($\Delta_k^{\phi} \simeq 1\,$),
unless some physical (model dependent) process drives them classical.
Among these processes, nonperturbative particle creation during 
preheating can drive to a classical level the sub-Hubble modes in the
resonance bands discussed in the previous
section~\cite{KT,PR,selfPE,selfFK}. 

In Fig.~\ref{deltamulti} we show instead the spectrum $\Delta_k^\chi$ in
the two-field model of preheating (\ref{twof}) for some values of $R$ 
(the largest time plotted corresponds to the moment at which Hartree
approximation  breaks down).  The case $R=3$ reproduces the results of the
one massless  field case upon identification of $\chi$ with the
fluctuations of the field  $\phi\,$.  For this value,
Fig.~\ref{deltamulti} shows the enhancement of  the modes in the resonance
band~(\ref{single}).  In the two-field case,  strong resonance occurs in
the parameter ranges described by  Eq.~(\ref{range}).  In the center of
the resonance bands, $R=2n^2$, small  $k$ modes ($\kappa~\lsim~0.5$) are
efficiently excited, while the modes  with $\kappa~\gsim~3$ are regarded
as quantum fluctuations  ($\Delta_k^{\chi} \simeq 1$) as long as $R$ is
not much larger than unity  (this evolution does not include rescattering
effects which become  important in the nonlinear regime).

In the next section we numerically evaluate the formula for the 
occupation numbers reported in Sec.~II. In the calculation we do not
include those modes which are not excited to a classical  level
($\Delta_k^{(i)} \simeq 1$). For modes which become classical, from the 
two decompositions~(\ref{esse}) and (\ref{deffik}) we get
\begin{equation}
{\tilde \Phi} \left( k \right) = \frac{V^{1/2}}{\sqrt{2}\,\left( 2 \, \pi 
\right)^{3/2}} \, \Phi_k \,\,.
\label{relpert}
\end{equation}

\subsection{Particle production}

We can combine eqs.~(\ref{npsi}) and (\ref{relpert}) to evaluate the
number density of produced fermions in terms  of the modes $\Phi_k$
computed in the previous section. Moreover, we assume the metric
perturbations to be statistically isotropic and homogeneous, so that
$\Phi_{\bf k} = \Phi_{\vert {\bf k} \vert}\,$. In this case, the final
particle density is given by
\begin{equation}
N_\psi \equiv \frac{N_2}{V} = \frac{1}{15 \, \left( 2 \, \pi \right)^4 \,
a^3} \int_0^\infty d p_0 \int_0^{p_0} d p \, p^6 \left|
\int_{\eta_i}^{\eta_f} d \eta \, {\rm e}^{i p_0 \eta} \, \Phi_p
\left( \eta \right) \right|^2 \,\,.
\label{npsi2}
\end{equation}
As discussed in the previous subsection, the integral over $p \equiv \vert
{\bf p} \vert$ must be restricted to the region where the perturbations
are classical. We have seen in section~\ref{secpert} that this typically
occurs for $p$ smaller than a cut-off value $p_{c}$ which approximately
corresponds to modes which never crossed the Hubble scale (or, in the
massless inflaton case, to the highest resonance band excited). In the
numerical computation of $N_2$ we also integrated $p_0$ up to $p_{c}\,$.
Higher frequencies are not expected to give a sizable contribution to
eq.~(\ref{npsi2}), since the rapidly oscillating phase 
${\rm e}^{ip_0\eta}$ averages to nearly zero the time integral. 

In the numerical calculations we take as the upper limit $\eta_f$ of the
time integral the moment in which backreaction effects start to shut off
the resonance. In the single self-coupling case this occurs at $x_f \equiv
\sqrt{\lambda}\varphi(t_{\rm co})\eta_f \simeq 500$, while in the
two-field case the precise value of $\eta_f$ is a function of the ratio
$R\,$. Note that this is just an approximate calculation, since the above
integral is defined rigorously only for the asymptotic future, $\eta_f \to
\infty$, with vanishing perturbations $\Phi_k \to 0\,$. That is, our
numerical calculation is approximately
\begin{equation}
n_{\psi} \equiv a^3 N_\psi \simeq \frac{1}{15 \, \left( 2 \, \pi 
\right)^4} \int_0^{p_c} d p_0 \int_0^{p_0} d p \, p^6 \left| 
\int_{\eta_i}^{\eta_f} d \eta \,
{\rm e}^{i p_0 \eta} \, \Phi_p \left( \eta \right) \right|^2 \,\,.
\label{npsi3}
\end{equation}

Before presenting the numerical results, we briefly consider the
occupation number of massive fermions produced by the  homogeneous FLRW
expansion, eq.~(\ref{N0}). For sufficiently small masses, the final
occupation number is still (approximately) given by~(\ref{npsi3}), since
the  Bogolyubov coefficient $\beta$ is very close to zero in this limit
(see  eq.~(\ref{bogo}) and (\ref{N2})). The term~(\ref{N0}), although
quadratic in $\beta\,$, is not suppressed by the small perturbations.
Thus, by solving the equation 
\begin{equation}
N_0 \left( m_\psi \right) = N_2 \left( m_\psi = 0  \right)
\end{equation} 
we can estimate up to which mass $m_\psi$ the
productions by the inhomogeneities is comparable with that
from the  homogeneous expansion.

The calculation of $N_0$ has been performed numerically in~\cite{igor} 
(see also references therein). For low fermionic masses, the gravitational
production from the homogeneous expansion mainly occurs when the Hubble
expansion rate equals $m_\psi\,$. After this time, the comoving fermion
number density is given by~\cite{igor}
\begin{equation}
n_\psi^{\rm hom} \equiv a^3 \, N_0 / V \simeq C_\alpha \, m_\psi^3 \left( 
\frac{H_0}{m_\psi} \right)^{3\,\alpha}\,\,.
\label{hom}
\end{equation}
In this expression, $H_0$ denotes the value of the Hubble rate at the
initial time $t_0\,$, where the scale factor $a \left( t_0 \right)$ is
normalized to unity. The parameter $\alpha$ is instead the exponent
appearing in the expansion law $a \left( t \right) \propto t^\alpha$ when
$H \left( t \right) = m_\psi\,$ (as is well known, the coherent inflaton 
oscillations give effective matter domination, $\alpha = 2/3\,$, in the 
$V=m^2\phi^2/2$ case, and effective radiation domination,  $\alpha
=1/2\,$, for $V=\lambda\phi^4/4$). For the first coefficient of
eq.~(\ref{hom}), the two values $C_{2/3} \simeq 3 \times 10^{-\,3}$ and
$C_{1/2} \simeq 10^{-\,2}$ have been numerically found~\cite{igor}.

\subsubsection{$V=\frac12m^2\phi^2$}

Let us first present the numerical results for the massive inflaton 
case.  The precise value deduced from the integral~(\ref{npsi3}) is sensitive to 
the cut-off $p_c$.  To discriminate which cut-off should be taken, we 
consider the quantity $\Delta_k^{\phi}$.  We found that $\Delta_k^{\phi}$ 
approaches unity around $k=10m$, so we chose $p_{c}=10 \,m \,$.  In this 
case, for the comoving occupation number (normalizing $a=1$ at $\phi=0.5 \, 
M_p$) we have numerically found
\begin{equation}
n_\psi \equiv a^3 N_{\psi} \simeq 3 \times 10^{-14}m^3
\label{nmass}
\end{equation}
(just to give an example on how the final result is sensitive to $p_c\,$, 
we found $n_{\psi} \simeq 1 \times 10^{-14}m^3$ for $p_{c}=5m$
and $n_{\psi}\simeq 1 \times 10^{-13}m^3$ for $p_{c}=20m$).

Equating (\ref{npsi3}) with (\ref{hom}), the mass $m_{\psi}$ below which 
the production from inhomogeneities dominates over the one from the 
homogeneous expansion is estimated as
\begin{equation}
m_{\psi}=\frac{n_{\psi}}{C_{2/3}H_0^2} \simeq 10^2~{\rm GeV}.
\label{mcompar}
\end{equation}

Of more physical interest is the fermionic abundance deduced 
from~(\ref{nmass}). By assuming an instantaneous inflaton decay into a 
thermal bath characterized by the reheat temperature $T_{\rm rh}$, 
the particle density divided by the entropy density, $s$, is given by
\begin{equation}
Y_\psi \left( T_{\rm rh} \right) \equiv \frac{N_\psi}{s}
\left( T_{\rm rh} \right) 
\simeq 10^{-\,29} \, \frac{T_{\rm rh}}{10^9 \, {\rm GeV}} \,.
\label{Yb}
\end{equation}
In the case of other light gravitational relics, such as  gravitini or
modulini, the primordial abundance is severely constrained by the
successful predictions of Big Bang nucleosynthesis. For masses of order
the electroweak breaking scale, the limit is about $Y~\lsim~10^{-\,13}$ --
$10^{-\,12}\,$~\cite{km}. In the present case, eq.~(\ref{Yb}) gives a much
smaller abundance.

\subsubsection{$V=\frac14 \lambda \phi^4$}

In the quartic case, both super-Hubble modes and the modes in the
interval~(\ref{single}) contribute to the particle production,
eq.~(\ref{npsi3}). From simple estimates (for example combining
eq.~(\ref{npsi3}) with the values for the perturbations shown in
Fig.~\ref{selfmet}), one can see that the main contribution to particle
production is given by the resonance band~(\ref{single}).

In calculating the total number densities of produced fermions due to
inhomogeneities, we take the cut-off value in momentum space at
$p_c=3$, over which $\Delta_k^{\phi}$ approaches unity. 
Then the final particle density reached at the
end of preheating is numerically found to be 
\begin{equation}
n_{\psi} \simeq 1 
\times 10^{-13}\,(\sqrt{\lambda}\phi_0)^3.
\label{nself}
\end{equation}
which is larger than in the massive inflaton case.

Proceeding as in the previous subsection we find that the production from 
metric perturbations dominates over that from the homogeneous expansion 
as long as the fermions have masses smaller than
\begin{equation}
m_{\psi}=\frac{1}{H_0} \left(\frac{n_{\psi}}{C_{1/2}} \right)^{2/3} \simeq 
10^5~{\rm GeV} \,\,.
\label{mcompar2}
\end{equation}
The final abundance is also greater than in the previous case (thus 
confirming that the integral~(\ref{npsi3}) is dominated by the modes in 
the resonance band),
\begin{equation}
Y_\psi \left( T_{\rm rh} \right) \simeq 10^{-\,22} \,\,.
\label{ab2}
\end{equation}
Nevertheless it is still too small to have any cosmological effect.

\subsubsection{$V=\frac14 \lambda \phi^4+\frac12g^2\phi^2\chi^2$}
Since this potential leads to an infinite number of resonance bands it 
is the case of most interest. 
In Fig.~\ref{npsif} our numerical results for fermionic production 
are presented as a function of $R \equiv g^2/\lambda\,$, for $R \leq 
10\,$. This is obtained by solving the linearized equation (\ref{Phim})
using the Hartree approximation for field perturbations.
 By comparison with the single field case ($g = 0$, here reproduced by the 
choice $R=3$) the two equations~(\ref{mcompar2}) and (\ref{ab2}) are 
generalized to
\begin{equation}
m_\psi \simeq 5 \times 10^{13} \: {\tilde n}_\psi^{2/3} \: {\rm GeV} 
\;\;\;,\;\;\; Y_\psi \left( T_{\rm rh} \right) \simeq 10^{-\,9} \: {\tilde 
n}_\psi \,\,,
\end{equation}
where ${\tilde n}_\psi \equiv n_\psi / (\sqrt{\lambda} \, \phi_0 )^3\,$.

The qualitative behavior of Fig.~\ref{npsif} is readily understood from the
structure of resonance~(\ref{range}). In the range of $R$ plotted, we have
maximal production when $R$ is in the center of the first resonance band,
in which case $n_\psi$ is found to be $\tilde{n}_\psi \simeq 5 \times
10^{-10}$, which is about $5000$ times larger than in
the single-field case. In the second resonance band, particle production is
slightly smaller than in the $R=2$ case. 

\begin{figure}
\begin{center}
\singlefig{10cm}{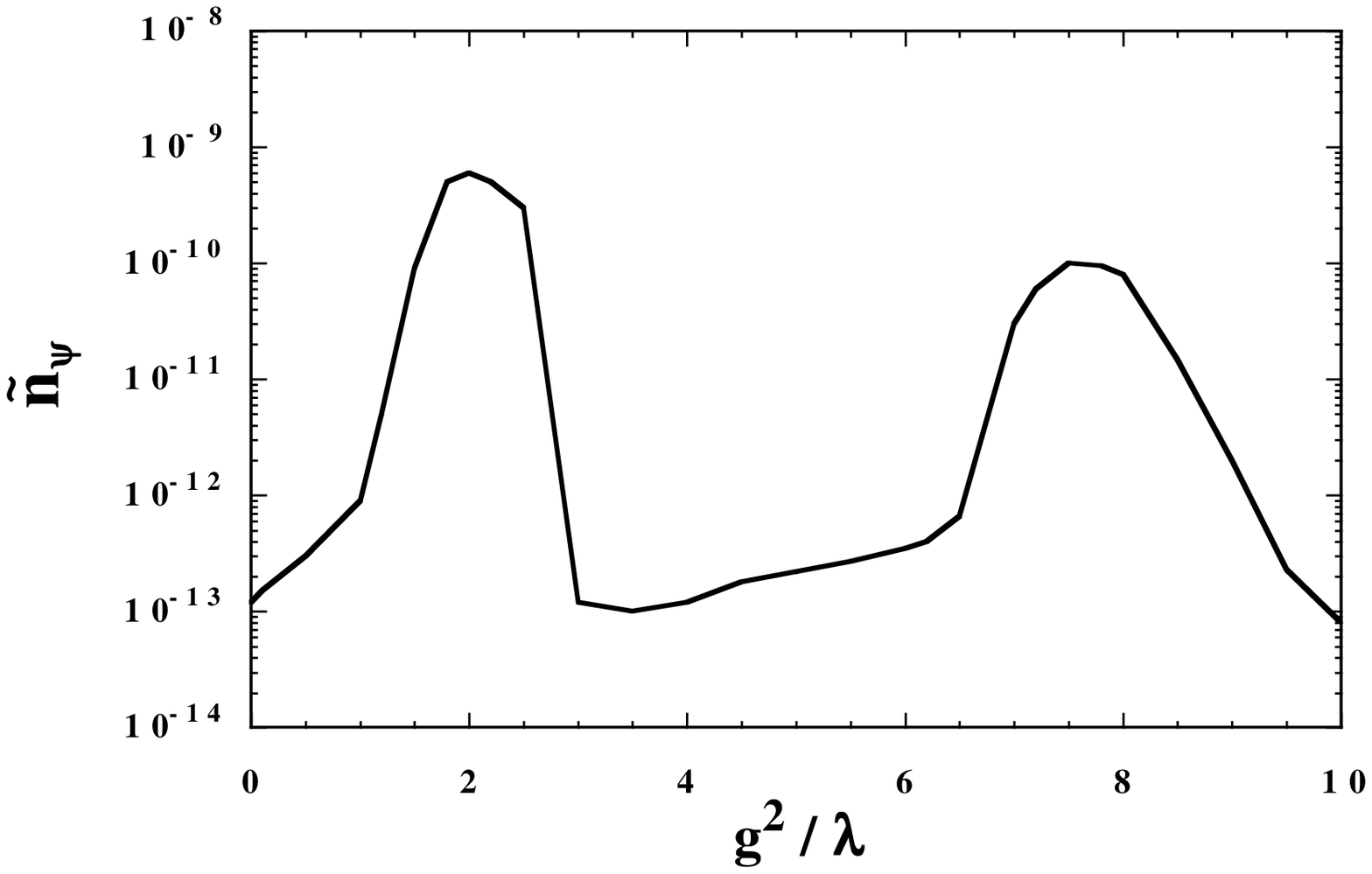}
\begin{figcaption}{npsif}{10cm}
The comoving number density, $\tilde{n}_{\psi} = n_{\psi}/
(\sqrt{\lambda}\phi_0)^3$, vs the ratio $g^2/\lambda$, for
$g^2/\lambda \le 10$. Note that this is based on the linear calculation 
using
eq.~(\ref{Phim}), which does not include the second order effect in field
perturbations. We take $\chi=10^{-3}M_p$ at $\phi=4M_p$. We see that
particle creation is enhanced in the super-Hubble 
resonance bands~(\ref{range}).
\end{figcaption}
\end{center}
\end{figure}

As $R$ increases, the inflationary suppression of $\chi$ becomes more and
more relevant, and particle production from first order 
perturbations is consequently reduced. Nevertheless, the second order
effect of field perturbations described in subsection~III B 3 can lead to 
the
excitation of metric perturbations on sub-Hubble scales. Just for
indicative purposes, we report a couple of our numerical results in this
range. For large $R$, the resonance bands cover only modes up
to~\cite{GKLS}
\begin{eqnarray}
\kappa~\lsim~\left(\frac{R}{2\pi^2}\right)^{1/4} \,\,.
\label{kappamax}
\end{eqnarray}
In the center of the 10-th instability band, $R=200$, one finds resonant
amplification up to $\kappa~\lsim~1.78$, as can be appreciated in
Fig.~\ref{deltamulti}. By implementing the analysis reported in 
subsection~III B 3
we find $\tilde{n}_{\psi} \simeq 6 \times 10^{-11}$.
As $R$ increases, higher momentum modes contribute  to the growth of
metric perturbations. For example, eq.~(\ref{kappamax}) gives
$\kappa~\lsim~4$ in the $R=5000$ case.  As a consequence, this 
enhances fermionic production. The comoving occupation number 
in this case is  found to be $\tilde{n}_{\psi} \simeq 1 \times 10^{-7} 
\,,\,\, Y_\psi \left( T_{\rm rh} \right)\simeq 10^{-16}\,$.

It is however important to mention that for $R \gg1$ the whole analysis
becomes very delicate. For large coupling $g\,$, the backreaction effect
of produced particles ends resonant amplification of field perturbations
earlier. In addition, it was found in Ref.~\cite{KT} that the final field
variances get smaller if rescattering of the $\delta\chi$ fluctuations is
taken into account in the rigid FLRW spacetime ($\Phi = 0$). On the
contrary, the effect of rescattering can lead to the excitation of
inflaton fluctuations through the amplification of $\delta\chi$
fluctuations. Since (contrary to the field $\chi$) the homogeneous
inflaton component is not suppressed during inflation, the gravitational
potential can then acquire a potentially large additional source (see
eq.~(\ref{Phim})). Finally, the anisotropic stress is expected to be
important in the nonlinear regime, in which case the relation $\Phi=\Psi$
no longer holds \cite{selfFK}. The study of these effects is certainly
worth separate detailed investigation.

\section{Conclusions}

In the present work we have discussed gravitational creation of fermions by
inhomogeneous scalar perturbations of a Friedmann-Lemaitre-Robertson-Walker
(FLRW) Universe. Massless fermions are conformally coupled to this
background, and thus are not generated by the homogeneous gravitational 
field. As a consequence, gravitational production of light 
particles is solely induced
by metric perturbations which break the conformal flatness of the 
background. 
Although the small size of perturbations on very large scales
indicated by CMB experiments seems to suggest a small production, a precise
computation is still worth while for light gravitational relics.

This is particularly true in scenarios in which
metric perturbations undergo parametric amplification during preheating at 
the end of inflation. To
examine this issue more quantitatively, we have studied some particularly
simple models of chaotic inflation. The simplest example is a 
single massive inflaton field ($V = m^2 \, \phi^2
/2$), where parametric resonance is actually absent. In this model, metric
perturbations are nearly constant during reheating for modes which
left the Hubble scale during inflation, whereas the modes deep inside the
Hubble scale exhibit adiabatic damping. As may be expected, particle 
production
is negligibly small in this situation. 

We then studied the massless self-coupled inflaton field 
($V = \lambda \, \phi^4 / 4$) case, in which
parametric resonance occurs for modes in a small band in momentum space 
near 
the Hubble scale.
As a consequence, particle production is enhanced. Moreover, for a
massless inflaton the background energy density decreases more quickly
with respect to the massive case, and this results in a larger abundance
at reheating of the particles generated by the inhomogeneities. Despite 
these two effects, the final result for the production is however still
well below the limit from nucleosynthesis.

More efficient production is expected when more fields are present, such 
as the model characterized by the potential
$V=\lambda \, \phi^4/4 + g^2\phi^2 \, \chi^2/2\,$. In this case, the
crucial parameter which determines the resonance bands is $R
\equiv g^2 / \lambda\,$. In particular, the longwavelength modes of the 
second
field $\chi$ are parametrically amplified whenever $R \simeq 2 \, n^2$
with integer $n\,$. At linear order in field perturbations, these modes
are coupled to metric perturbations through a term proportional to 
$\dot{\chi} \delta \chi_k\,$. For relatively small $R\,$, metric 
perturbations
are amplified by this term, and particle production is strongly
enhanced. 

For $R > 10$  the large $\chi$ effective mass 
($m_\chi = g \, \langle \phi \rangle \propto R^{1/2}$) causes the $\chi$ 
field to be exponentialy suppressed during inflation.
As a consequence, for large $R$ metric perturbations mainly increase due to
second order field fluctuations~\cite{liddle}. Particle
production may be expected to become more significant at very large $R\,$.
This is indeed suggested by our numerical results, although in the cases
considered the final abundance is still below the nucleosynthesis
bounds. 

Our numerical simulations include the backreaction of the fluctuations on
the background evolution in the Hartree approximation. Rescattering
effects -- mode-mode coupling between fields -- are instead neglected.
This may considerably affect the final abundances, particularly in the
large $R$ limit where the mode coupling becomes particularly important. 
Strong rescattering effects imply an earlier end to preheating, which may
decrease the final values of field and metric perturbations. On
the other hand, rescattering leads to amplification of modes outside the
resonance bands. In particular, inflaton field fluctuations can be
significantly amplified. Since (contrary to the field $\chi$) the 
homogeneous 
inflaton component is not suppressed during inflation, metric
fluctuations may be strongly enhanced by the source term proportional
to $\dot{\phi} \delta\phi_k$ (analogous to the $\dot{\chi} \delta 
\chi_k$ term considered above). This  would be reflected in more
particle creation.

Stronger production from inhomogeneities may also occur in different
contexts. In the multi-field inflationary scenario, the gravitational
potential can exhibit nonadiabatic growth during inflation \cite{SY,GW},
especially when the second field $\chi$ has a negative nonminimal coupling
to the Ricci scalar~\cite{TY,STY}. In addition, it was shown that the 
growth
of metric perturbations can be present even in the single field case in
the context of non-slow roll inflation with potential gap~\cite{LS}, and
during graceful exit in string cosmology~\cite{KS}. It may be
interesting to study gravitational particle production in these contexts 
along the lines presented in this work.

%
\section*{ACKNOWLEDGEMENTS}
The authors  thank J\"urgen Baacke and 
Fermin Viniegra for contributions at early stages of this project
and Robert H. Brandenberger for comments on the draft.
ST also thanks Alexei A. Starobinsky and Jun'ichi Yokoyama
for useful discussions. 
ST is thankful for financial support from 
the JSPS (No. 04942). The work of MP is supported by the European 
Commission RTN programmes HPRN-CT-2000-00148 and 00152. MP thanks Lev 
Kofman for some useful discussions. He also thanks the Canadian Institute 
for Theoretical Astrophysics of Toronto and the Astrophysics group of 
Fermilab for their friendly hospitality during the early stages of this 
work.


\end{document}